\documentclass{PoS}

\title{The fate of $U_A(1)$ and topological features of QCD at finite temperature}

\ShortTitle{The fate of $U_A(1)$ and topological features of QCD at finite temperature}

\author{\speaker{Sayantan Sharma}%
         \thanks{Current Address: Helmoltz Institute, Johannes Gutenberg University, Mainz}\\
        [for HotQCD Collaboration]\\
        Brookhaven National Laboratory\\
        Upton NY 11973\\
        E-mail: \email{sayantans@bnl.gov}}

\abstract{The nature of chiral phase transition for QCD with two light quark flavors is not yet completely resolved. This is primarily because
 one has to understand whether or not the anomalous U(1) symmetry in the flavor sector is effectively restored along with the chiral symmetry.
Since the physics near the chiral phase transition is essentially non-perturbative, we employ first principles lattice techniques to address this issue. 
We use overlap fermions, which have exact chiral symmetry  on the lattice, to probe the anomalous U(1) symmetry violation of 2+1 flavor dynamical QCD 
configurations with domain wall fermions. The latter also optimally preserves chiral and flavor symmetries on the lattice. We observe that the anomalous 
U(1) is not effectively  restored in the chiral crossover region.  We perform a systematic study of the finite size and cut-off effects since the signals 
of U(1) violation are sensitive to it.  For the same reasons we also compare our results from the continuum extrapolated results of the QCD Dirac spectrum 
obtained from a different lattice discretization called Highly Improved Staggered Quarks. Our studies also provide a glimpse of the microscopic topological 
structures of the QCD medium that are responsible for the strongly interacting nature of the quark gluon plasma phase and related to the physics of 
confinement and chiral symmetry breaking.
}

\FullConference{Critical Point and Onset of Deconfinement - CPOD2017\\
		7-11 August, 2017\\
		The Wang Center, Stony Brook University, Stony Brook, NY}

\begin{document}

\section{Introduction}
Symmetries determine the order parameter across a phase transition. However for strongly interacting gauge theory described by Quantum Chromodynamics (QCD) the $U_A(1)$ symmetry, 
even though anomalous, could affect the nature of phase transition~\cite{pw} for $N_f=2$ massless quark flavors. Depending on whether $U_A(1)$ is effectively 
restored at the chiral transition temperature, the order of the phase transition and its universality class or both can change. All these arguments are based on perturbative 
renormalization group studies~\cite{pw,bpv} or conformal
bootstrap analysis~\cite{naka} of a model quantum field theory with the same symmetries as QCD. The magnitude of the $U_A(1)$ breaking term is just a parameter in such calculations.
How severely $U_A(1)$ is broken can be answered only non-perturbatively. Lattice gauge theory is one such suitable tool to study this problem and in spite of the challenges has led to 
quite a few independent studies \cite{shailesh,dw12,cossu,hiroshi,dw14,viktor,viktorqm, brandt,jlqcd17, jlqcd171}.  However there is still no consensus 
within the lattice community on whether $U_A(1)$ is badly broken or effectively restored at the chiral crossover transition temperature. 
The following questions still remains to be answered satisfactorily before a conclusive understanding of the problem can be made. 
\begin{itemize}
 \item The $U_A(1)$ breaking is believed to be due to the low-lying eigenvalues of the QCD Dirac operator. Do we understand the infrared region of the 
 eigenvalue spectrum without being affected by lattice discretization and finite volume effects?
 \item How close is QCD with 2 light but finite quark mass to the $N_f=2$ massless scenario?
 \item What are the microscopic topological objects that conspire to break $U_A(1)$.
 \item At high temperatures it is believed that the dilute instanton gas model could be a viable microscopic description for the low-lying spectra of the QCD 
 Dirac operator. Is it known from first principles if this model mimics QCD at high temperature? If yes then how ``high'' is this temperature?
 \item If indeed such temperatures are not asymptotically high what are its consequences especially in determining the QCD axion mass or alternatively the 
 axion decay constant. 
\end{itemize}
In the subsequent sections we will provide our results towards answering these questions. Before we proceed we briefly mention the technical details of our 
calculations for completeness.

\section{Numerical details}
We compare the eigenvalue spectrum of the QCD Dirac operator with different lattice discretizations to study carefully the discretization effects and 
ultimately try to understand the fate of $U_A(1)$ at finite temperature. We study two different fermion discretizations on the lattice, Highly Improved 
Staggered quark (HISQ) and  M\"{o}bius domain wall fermions. The QCD configurations with HISQ discretization were generated on $N_s^3\times N_\tau$ 
lattices where $N_s=4 N_\tau$ and $N_\tau$ was chosen to be $8,10,12,16$ respectively, taken from Ref.~\cite{hotqcdeos}. The quark masses are physical 
which gives a Goldstone pion mass of $140$ MeV for temperatures $T<200$ MeV for most of our ensembles. For higher temperatures the light quark masses are 
chosen to be slightly heavier than physical values giving a corresponding Goldstone pion mass of $160$ MeV. The lattice extent along the spatial directions 
is chosen such that it is atleast four times the pion Compton wavelength to ensure the volumes are large enough. Typical lattice spacings for the our $N_\tau=6$ 
lattices are of the order of $0.07$ fm. 
The other set of configurations generated with M\"{o}bius domain wall fermion discretization has Iwasaki gauge action with a dislocation suppressing determinant.
They have been taken from Ref.~\cite{dw14}. The pion masses for these sets are $135$ and $200$ MeV respectively and the lattice size is $32^3\times 8$.  The low-lying eigenvalues for these ensembles 
were measured with the overlap operator which has exact chiral symmetry and an index theorem on a finite lattice. The overlap operator was realized by calculating the sign function 
exactly with the eigenvalues of $D_W^\dagger D_W$ for low modes and representing the higher modes with a Zolotarev Rational function with $15$ coefficients. The sign function 
was measured to a precision of about $10^{-10}$ and the violation of the Ginsparg Wilson relation was of the same order of magnitude for each configuration. We chose the domain 
wall height appearing in the overlap operator $M=1.8$, which gave the best approximation to the sign function and satisfied the Ginsparg Wilson relation with the best precision. 
On each configuration $25$-$50$ eigenvalues of $D_{ov}^\dagger D_{ov}$ were measured using the Kalkreuter-Simma Ritz algorithm~\cite{ks}. We used about $100-150$ configurations at each 
temperature and for each value of the pion mass. In the subsequent sections, the temperatures quoted in our study are always written in units of $T_c$, where 
$T_c=154$ MeV being the chiral crossover transition temperature.
 
\section{The eigenvalue spectrum of QCD Dirac operator and the fate of $U_A(1)$}
\begin{figure}
\includegraphics[width=0.5\textwidth]{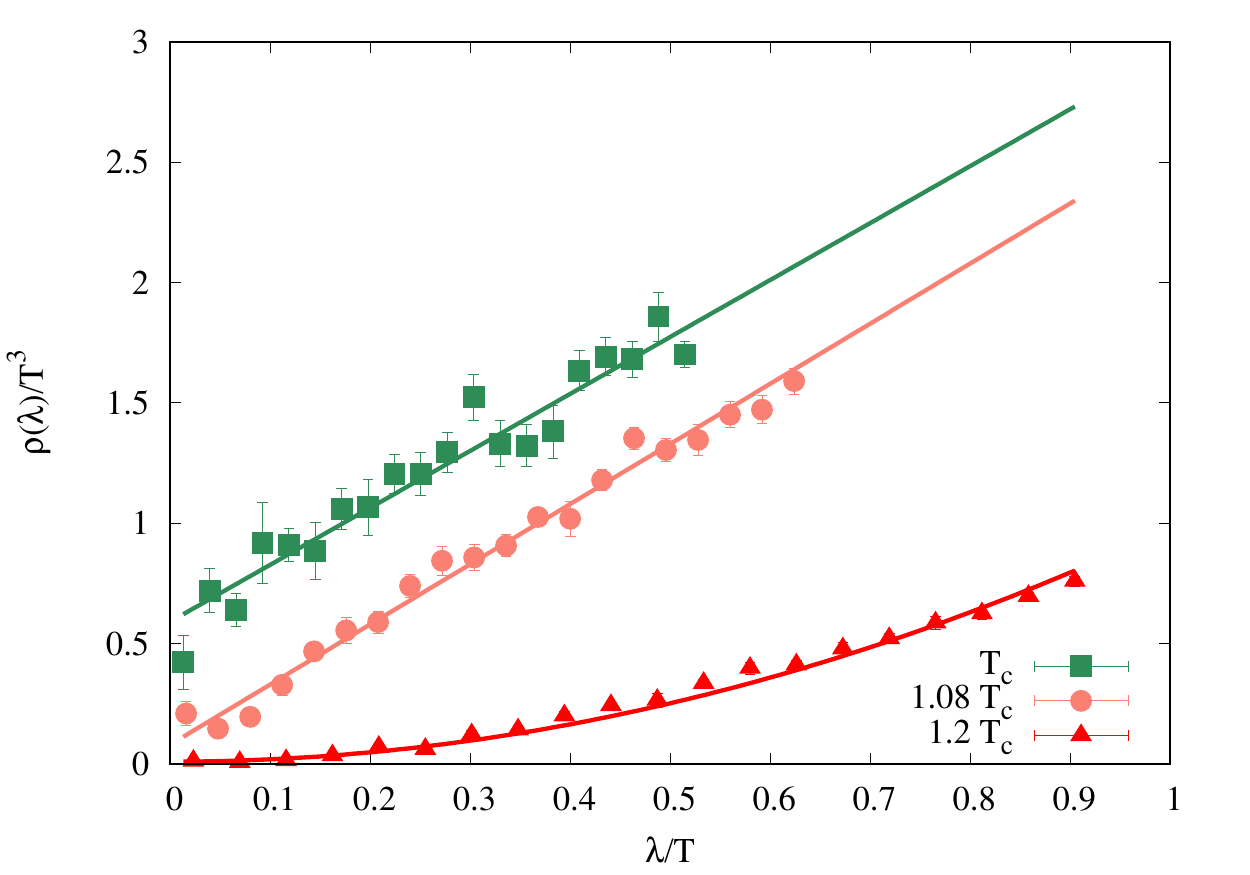}
\includegraphics[width=0.5\textwidth]{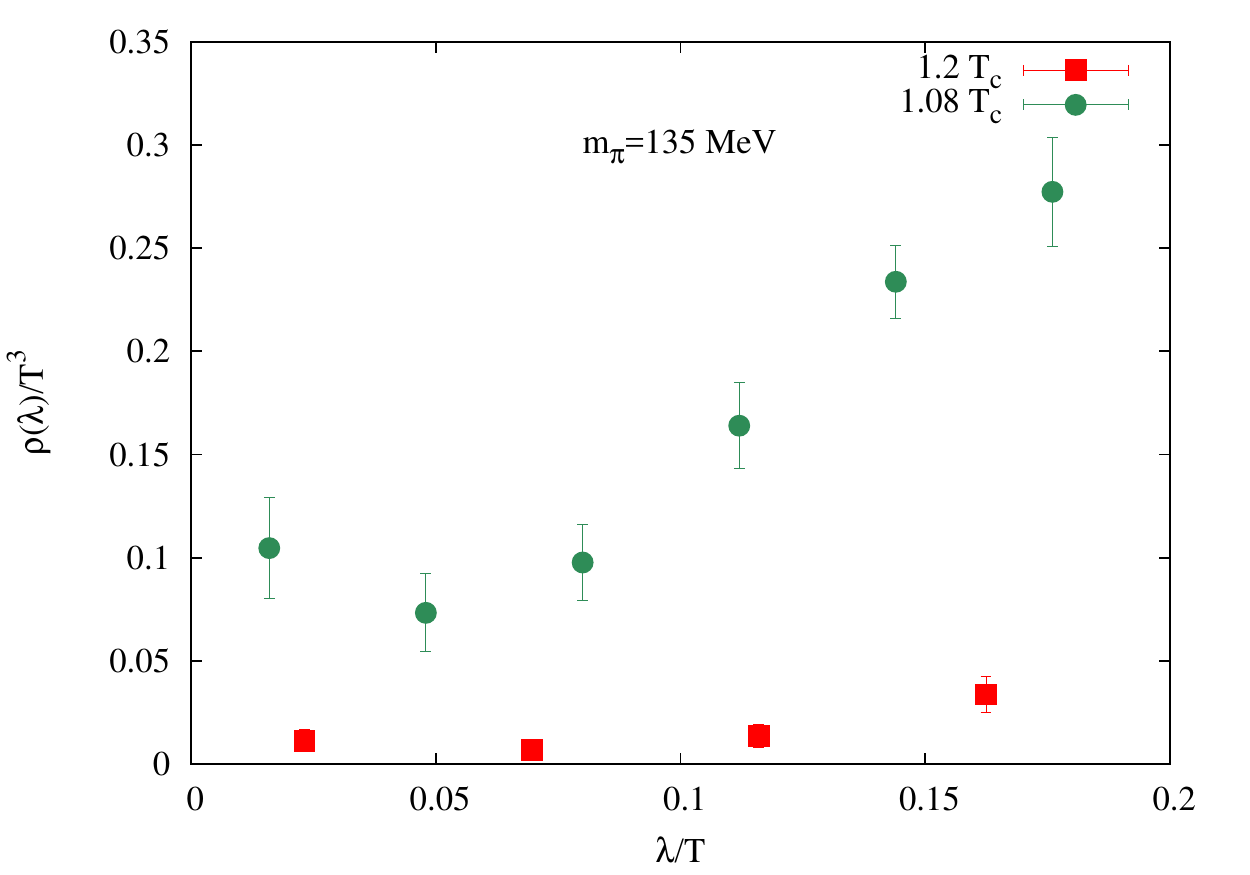}
\caption{The eigenvalue distribution of  M\"{o}bius domain wall fermions for different temperatures (left panel) and the near-zero mode peak zoomed in (right panel) for sea-quark masses 
measured with the overlap operator. }
\label{fig:eigdw}
\end{figure}

The eigen spectrum of the domain wall fermion ensembles measured with overlap Dirac operator, at three different temperatures $T_c, 1.08 ~T_c$ and $1.2~ T_c$ and for physical 
quark masses are shown in the left panel of Figure \ref{fig:eigdw}. These are the first 25 eigenvalues so we are in the deep infra-red sector of the eigenvalue density. As the 
temperature increases the near-zero peak (non-analytic) modes becomes distinguishable from the bulk spectrum which is denoted as an analytic function in $\lambda$.
To quantify the relative importance of the analytic and the non-analytic parts of the eigenvalue spectrum, we use the ansatz $\rho(\lambda)=\frac{A}{\lambda^2+\rho^2}+c |\lambda|^\gamma$ 
to describe the infrared part of the eigenvalue spectrum. The fit results for $\gamma$, which characterize the leading order rise of the bulk modes in the deep infrared, has a significant temperature dependence. 
Near $T_c$ the exponent $\gamma\sim 1$ consistent with the chiral perturbation theory estimates, which changes to $\gamma\sim 2$ at $1.2~ T_c$ consistent with our earlier observation from 
the HISQ eigen spectrum measured with the overlap operator~\cite{viktor}. Moreover $\gamma$ is found to be insensitive to quark mass so it can be expected that its temperature dependence 
will not change in the chiral limit as well~\cite{viktor,viktorqm}. Since we are in the deep-infrared part we are not sensitive to the $\vert\lambda^3\vert$ dependence expected from perturbation 
theory in the ultra-violet part of the eigenvalue spectrum. The analytic part of the eigenvalue spectrum of the QCD Dirac matrix at finite temperature has been recently studied in 
detail~\cite{aoki}. It was shown that when chiral symmetry is restored, using chiral Ward identities for upto $3$-point correlation functions, the leading order analytic part of the 
eigen spectrum goes as $|\lambda|^3$. It was further shown that the $U_A(1)$ breaking effects are invisible in the mesonic correlators in the scalar and pseudo-scalar sectors in up 
to $6$-point correlation functions~\cite{aoki} with this analytic dependence. 
Our results show that the leading $\lambda^3$ behavior of the infrared part of the eigenvalue spectrum will only set in after $1.2~T_c$, which also implies that $U_A(1)$ breaking 
effects due to the analytic part of the eigenvalue spectrum survive even at $1.2 ~T_c$. We next zoom into the tiny non-analytic part of the eigen spectrum as shown in the right panel of 
Figure \ref{fig:eigdw}.  There is a peak at $1.08~T_c$ for $\lambda/T<0.05$ which reduces significantly at $1.2~T_c$.

\begin{figure}[h] 
\includegraphics[width=0.5\textwidth]{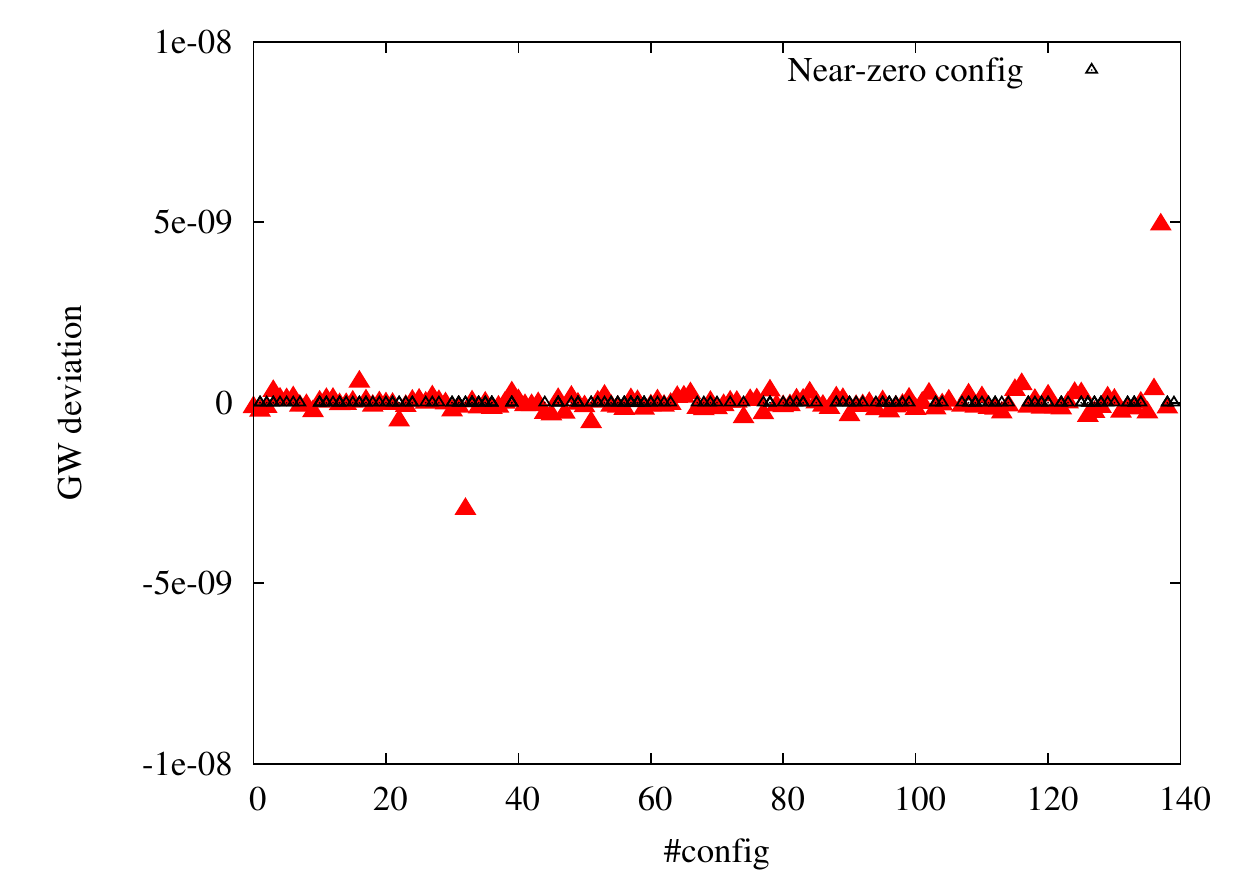}
\includegraphics[width=0.5\textwidth]{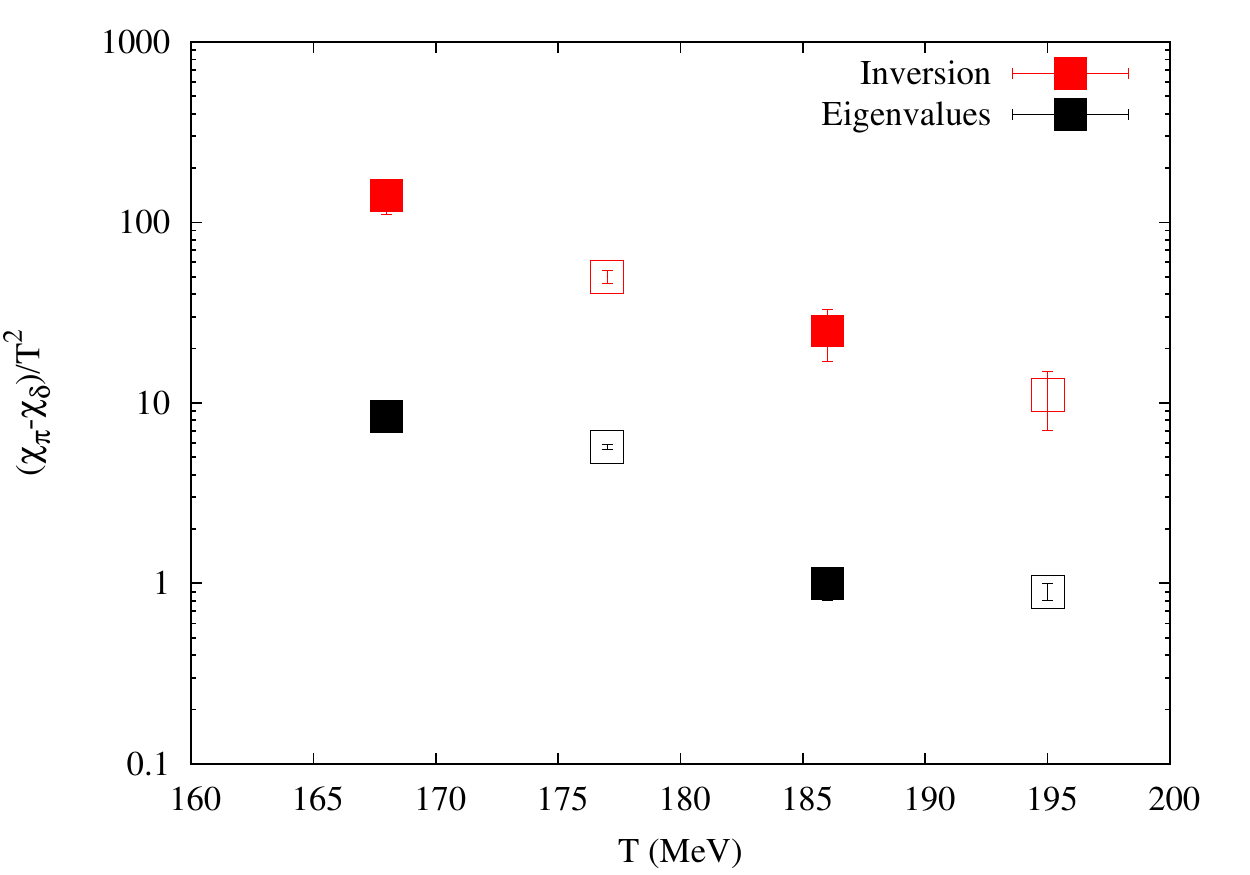}
\caption{The red points denote violation of the Ginsparg Wilson relation for the overlap operator for each M\"obius domain wall fermion configuration 
at $1.2 ~T_c$ for physical quark mass (left panel) and the black points mark those configurations which have 
near-zero modes. In the right panel the temperature dependence of $(\chi_\pi-\chi_\delta)/T^2$ is shown for different light quark masses, $m_l/m_s\sim 1/12$ 
(with open symbol) and $m_l/m_s\sim 1/27$ (solid) obtained from exact inversion of the Dirac operator using stochastic vectors (red) versus that are calculated 
from the first 25 eigenvalues (black).  }
\label{fig:gwdw}
\end{figure}

It has been earlier observed that on small lattice volumes, the non-analytic part of the eigenvalue spectrum of M\"{o}bius domain wall configurations can arise primarily due to violations of 
the Ginsparg Wilson relation~\cite{cossu2}. We compare the Ginsparg Wilson relation violation in the configurations which have near-zero modes at $1.2 ~T_c$, marked as black 
triangles in the left panel of Figure \ref{fig:gwdw} to the average magnitude of violation marked as red triangles. We do not observe any direct correlation between the violation of Ginsparg Wilson 
relation and the occurrence of near-zero modes. 
We next study the $U_A(1)$ violation as a function of temperature in the domain wall fermion ensembles due to the infra-red part of the eigenvalue spectrum 
consisting of the earlier measured 25 lowest eigenvalues. The degeneracy of the two-point correlation functions in the pion and the delta channel is a possible signature 
for the effective restoration of $U_A(1)$~\cite{shuryak}. The difference of the integrated correlators of pion and delta meson correlators, $\chi_\pi-\chi_\delta$ normalized by $T^2$, 
calculated from the eigenvalues shown in Figure ~\ref{fig:eigdw} for different light quark masses, $m_l$ are shown in the right panel of Figure \ref{fig:gwdw}. These 
values are compared with the values obtained on the same configurations from the exact inversion of the domain wall operator~\cite{dw14}. For comparison, we have tuned the overlap 
valence quark mass $m_l$ to the domain wall sea-quark mass. This was done by choosing a $m_l$ for constructing the RG invariant quantity 
$\frac{m_s\langle\bar\psi\psi\rangle_l-m_l\langle\bar\psi\psi\rangle_s}{T^4}$ from the first 25  eigenvalues measured by the overlap operator 
keeping $m_s/m_l\sim 12, 27$ respectively. Then the $m_l$ was tuned by setting the value of this quantity equal to the value obtained in ~\cite{dw14} by inverting 
the domain wall operator on the same configurations. It is clearly evident that the contribution of the lowest 25 eigenvalues out of a million corresponding to our 
lattice size, to the $U_A(1)$ breaking observable is significant almost $\sim 10\%$.

\begin{figure}[h]
\includegraphics[width=0.5\textwidth]{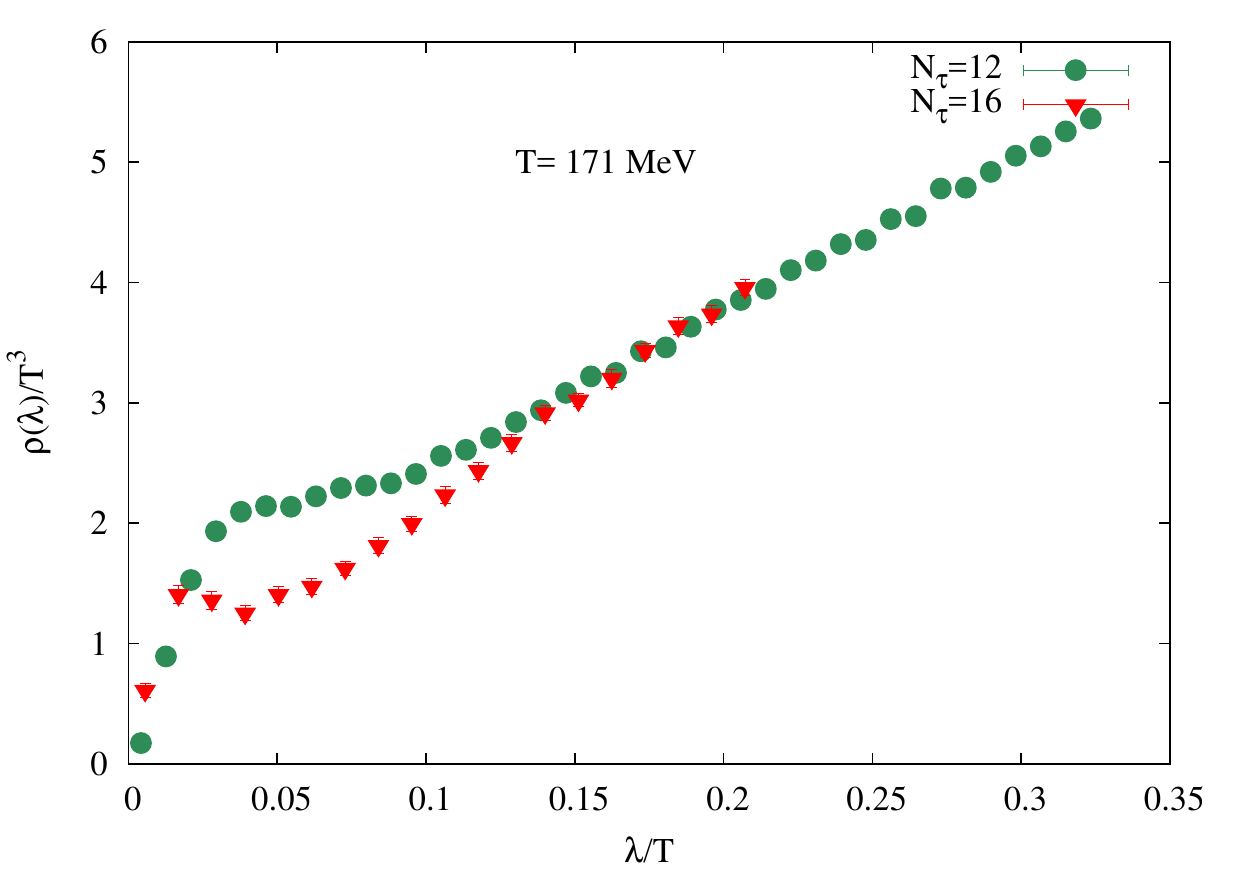}
\includegraphics[width=0.5\textwidth]{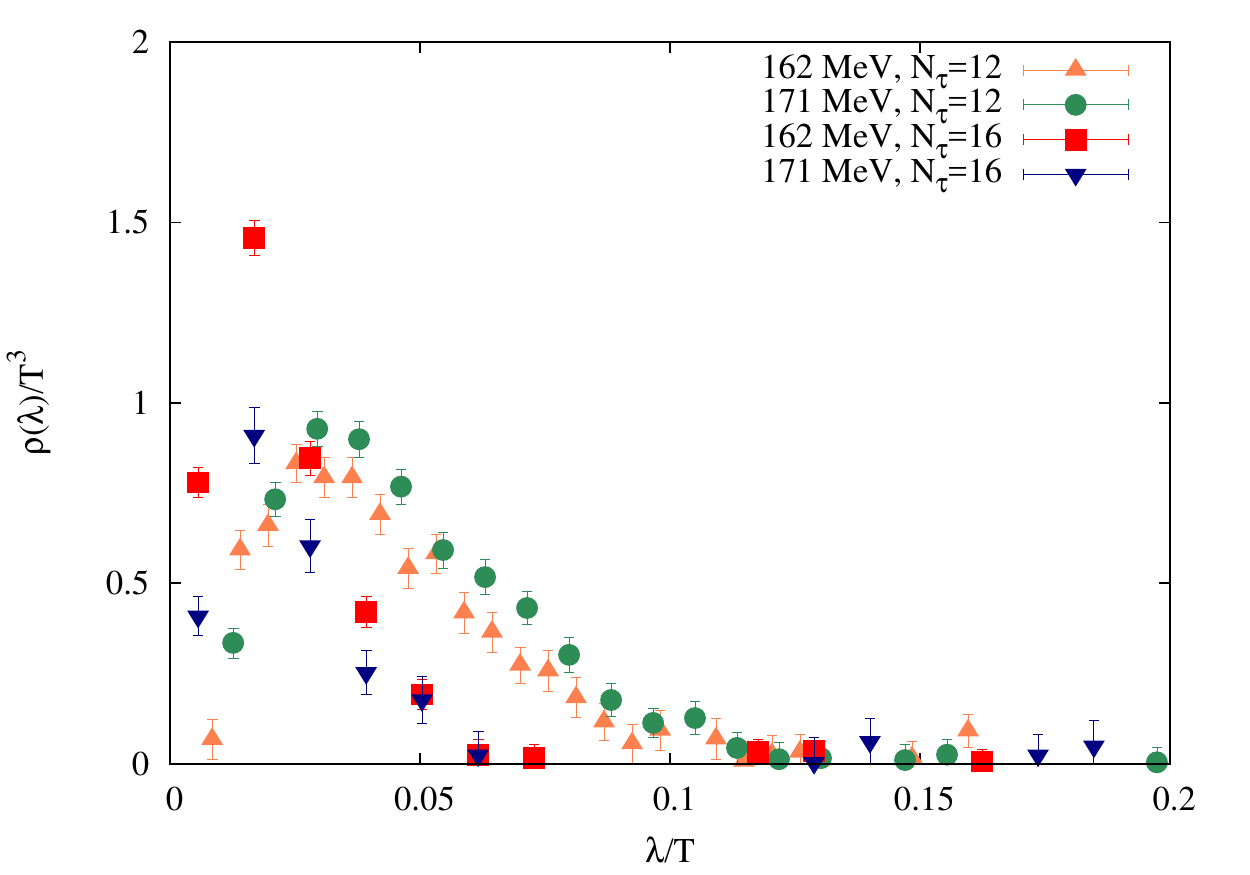}
\caption{The eigenvalue distribution of the  HISQ configurations for $N_\tau=12, 16$ lattices at $T \sim 171$ MeV (left panel) and near-zero peak region zoomed in (right panel) at the same 
temperature. }
\label{fig:dwcomp}
\end{figure}

We want to compare if our observations about the eigenvalue spectrum is independent of the lattice discretization. For this purpose we study the eigenvalue 
spectrum of the QCD Dirac operator with the staggered (HISQ) discretization. It has been already observed that the infrared eigenvalue spectrum has many interesting 
features for relatively coarser $N_\tau=8$ lattices~\cite{hiroshi}. We studied the spectrum for finer lattices by changing from $N_\tau=12$ to $N_\tau=16$, results of 
which are compiled in Figure \ref{fig:dwcomp}. We observe that as the lattice is made finer the non-analytic part gets distinguishable from the bulk and the general characteristics resemble 
the eigenvalue spectrum of the domain-wall fermions which respect chiral symmetries to be a better extent on the lattice. The analytic part of the HISQ eigenvalue spectrum 
is still characterized by an exponent $\gamma\sim1$ at temperature $1.1~T_c$, which shows that the analytic part is quite robust irrespective of the lattice discretization we choose. This 
part explicitly breaks $U_A(1)$. Using the fit ansatz we used to characterize the eigenvalue spectrum of the domain wall fermions earlier we can extract the peak 
of the eigen spectrum in the deep infrared. This extracted peak size is shown as a function of the lattice spacing in the right panel of Figure \ref{fig:dwcomp}. As one goes to 
a finer lattice the peak height increases and the width become sharper which gives us the confidence that this non-analytic part of the eigenvalue spectrum will survive 
in the continuum limit near $T_c$. This will also contribute to  $U_A(1)$ breaking explicitly. 

\begin{figure}[h]
\begin{center}
\includegraphics[width=0.3\textwidth]{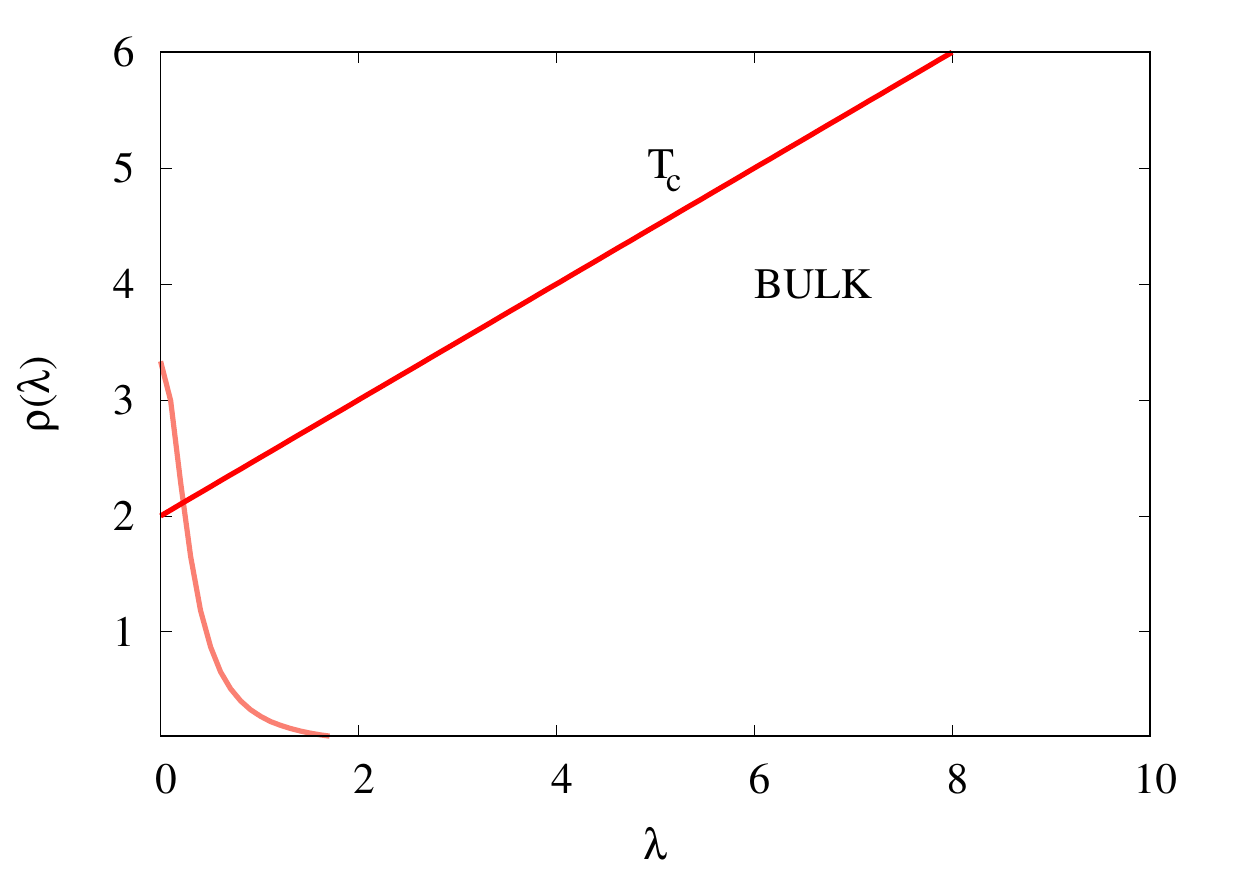}
\includegraphics[width=0.3\textwidth]{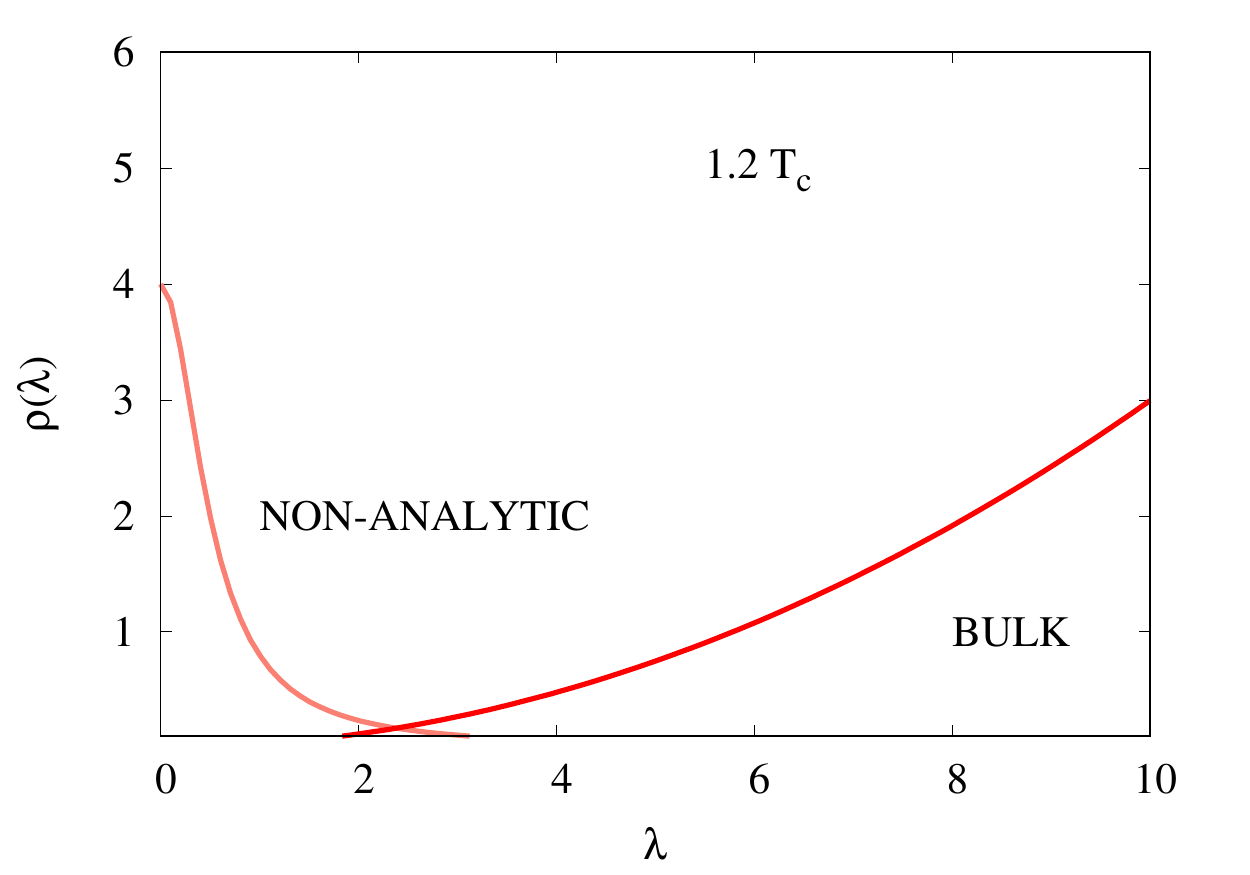}
\includegraphics[width=0.3\textwidth]{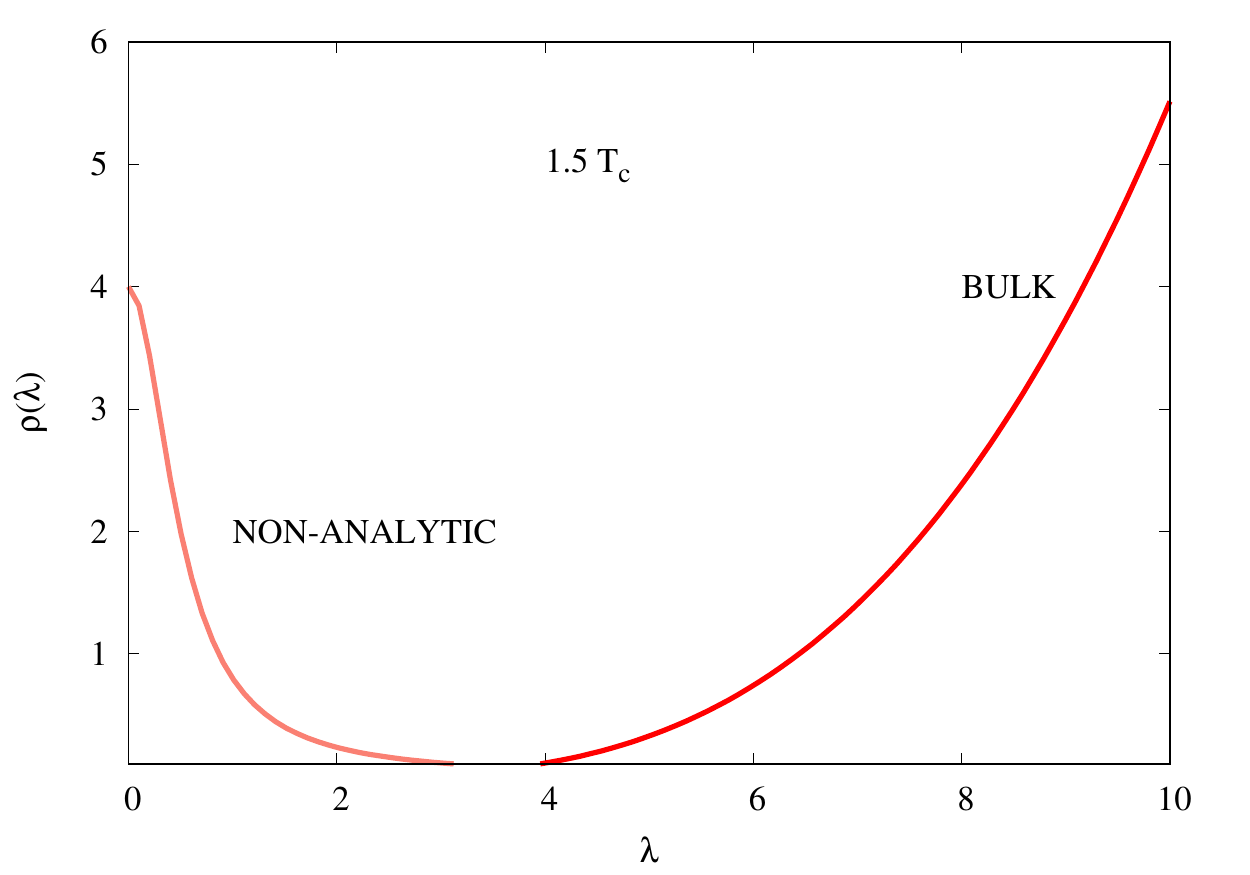}
\caption{The cartoon for the near-zero eigenvalue distribution in QCD as a function of temperature.}
\label{fig:eigsummary}
\end{center}
\end{figure}

We summarize our results here along with our earlier findings in ~\cite{viktor} in form of a cartoon plot of eigenvalue spectrum of QCD in Figure \ref{fig:eigsummary}.  The infrared part of the eigenvalue 
spectrum consists of an analytic part and a non-analytic peak. At $T_c$ these two parts superpose strongly making the near-zero peak indistinguishable from the bulk. As 
the temperature increases the bulk separates from the near-zero peak, making it identifiable. How far in temperature this peak survives is still debatable. Previous study with domain wall fermions
~\cite{dw12,dw14} and using HISQ discretization for quarks~\cite{viktor} show that it may survive upto $1.5 ~T_c$ whereas other groups studying the eigenvalue spectrum of the overlap 
and reweighted domain wall fermions~\cite{cossu} claim these may vanish immediately above $T_c$. However the analytic part survives and has a very characteristic 
temperature dependence quantified by the exponent $\gamma$ that gives its leading order rise. This part strongly contributes to $U_A(1)$ breaking till upto $1.5~T_c$ when the 
characteristic exponent $\gamma\sim 3$ and subsequently no longer contributes to $U_A(1)$ breaking correlators. The tiny peak which we still observed earlier at $1.5~ T_c$ ~\cite{viktor}
can be be interpreted as arising due to a dilute gas of instantons , i.e, associated with an instanton-antiinstanton pair. At lower temperatures the existence of the peak 
can be motivated from a quasi-instanton picture~\cite{yama}. Our study~\cite{viktor} suggests that the average instanton density is $0.147(7) fm^{-4}$ at $1.5~ T_c $, hence the 
dilute instanton gas description may already describe the QCD medium at $1.5~ T_c$. Similar features in eigen spectrum are also observed with stout-smeared staggered quarks at 
$T>T_c$ on even finer lattices~\cite{ivan}.

\section{The topological susceptibility in QCD from $T_c$ to high temperatures}
Another measure of the topological fluctuations in QCD is the topological susceptibility. It is defined as $\chi_t=\frac{T\langle Q^2 \rangle}{V}$ in the
thermodynamic limit where $Q^2$ measures the variation of the number of zero modes of the Dirac operator. Since tunneling between different topological 
sectors becomes rarer as one increases the temperature, $\chi_t$ is extremely tiny as temperature increases and has strong sensitivity to the lattice cut-off 
effects. It is thus important to properly perform the continuum limit details of which are described in ~\cite{peter}. Topological susceptibility also determines the mass of 
the QCD axions when the axion feels the tilt of the potential induced by  QCD and starts a slow-roll near its minima. This condition is met at a temperature $T$
when $m_a(T)=3~H$, where $H$ is the Hubble parameter. If indeed axions are probable dark-matter candidates and a fraction or even the total dark matter density 
in the present universe is due to the axions, then determination of $\chi_t$ indirectly constraints the decay constant of the axions.

\begin{figure}[h]
\includegraphics[width=0.5\textwidth]{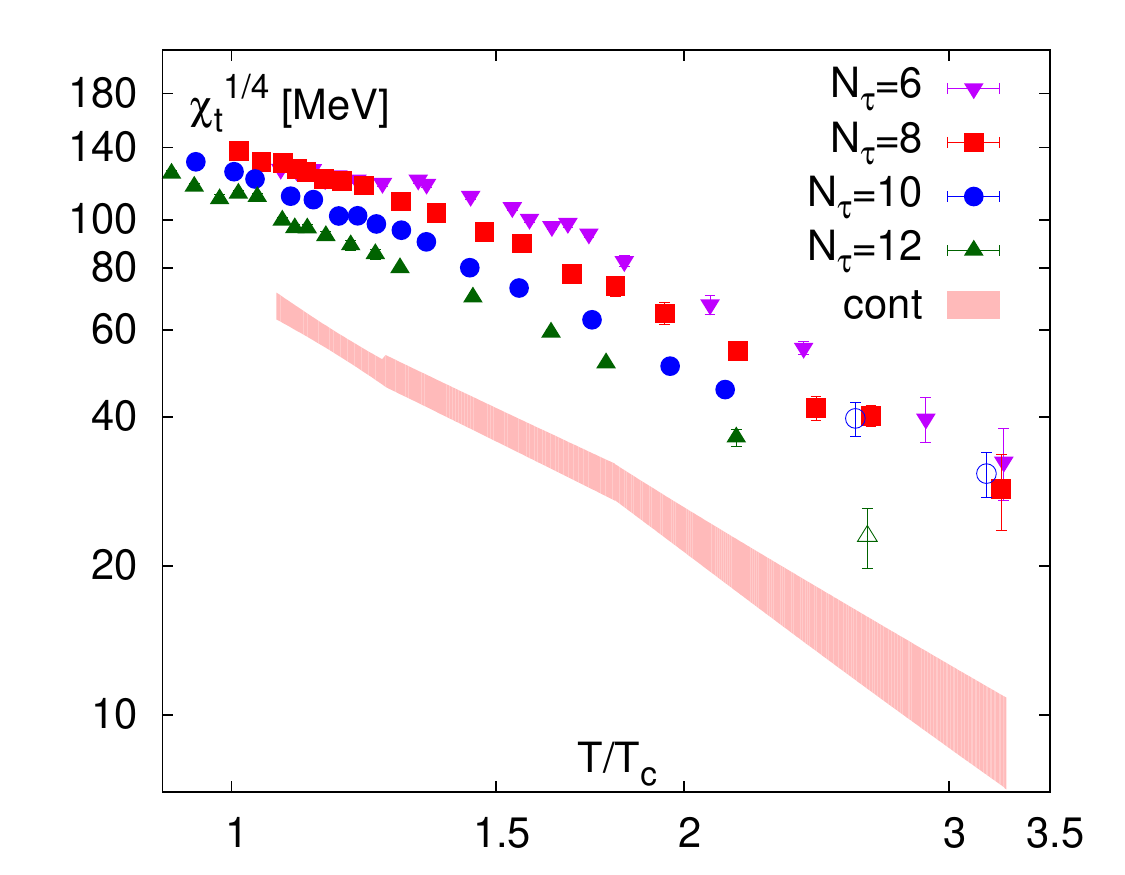}
\includegraphics[width=0.5\textwidth]{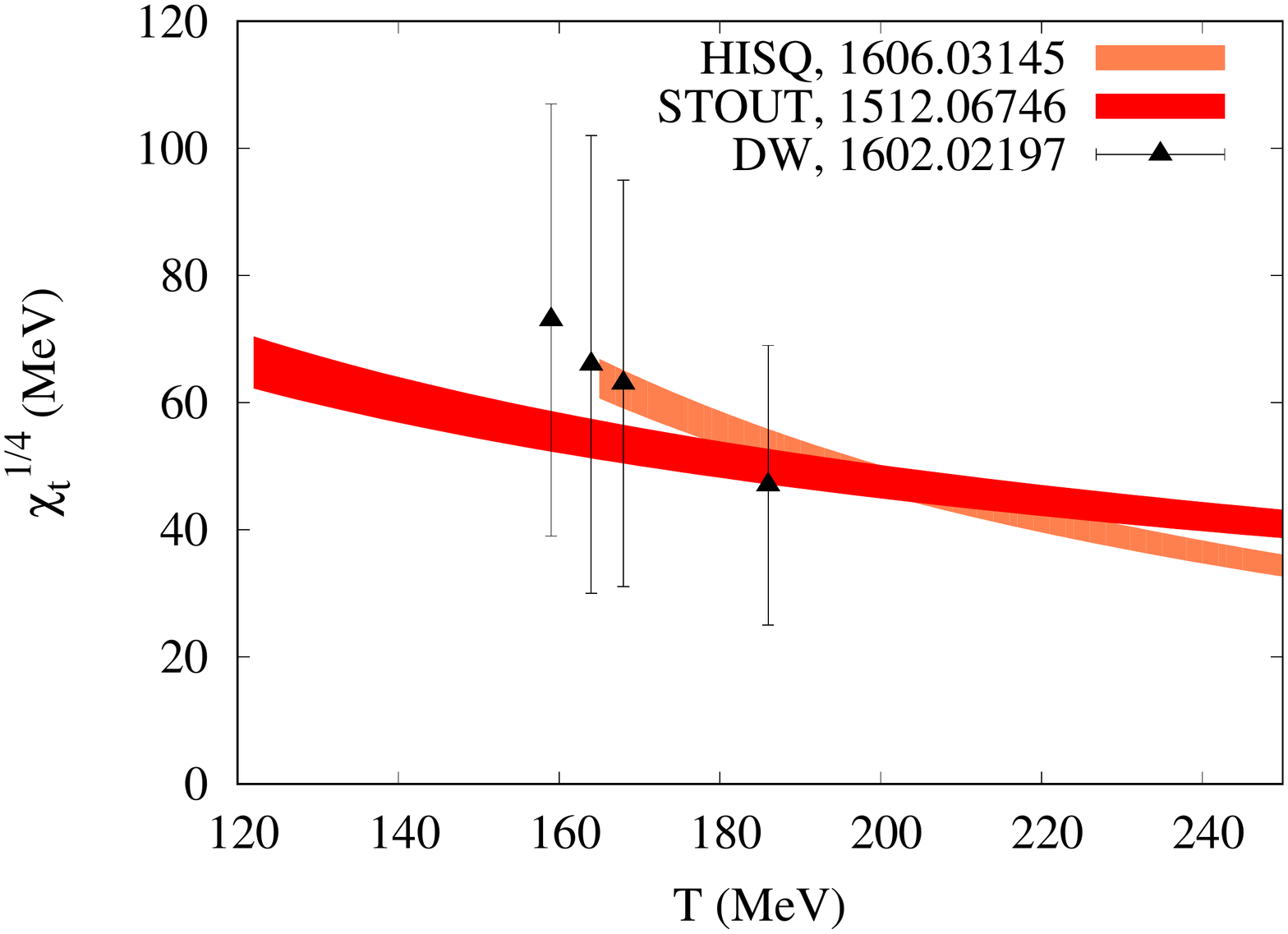}
\caption{The continuum extrapolated results for the topological susceptibility in QCD using HISQ discretization measured with the standard 
gluonic definition (left panel) from Ref.~\cite{peter}. In the right panel the results for the continuum extrapolated values of $\chi_t$ using different staggered 
discretization is compared with the results using domain wall fermions.}
\label{fig:topsusc}
\end{figure}

In the left hand panel of Figure \ref{fig:topsusc}, our results for the topological susceptibility in QCD is shown as a function of temperature 
after performing continuum extrapolation from Ref. ~\cite{peter}. The QCD ensembles were generated using the HISQ discretization for the fermions. 
The gauge configurations were systematically smeared using Wilson flow to remove the UV noise and then  the topological 
susceptibility was measured using the gluonic definition $\chi_t=\int d^4 x \langle F \tilde F(x)  F \tilde F(0)\rangle$. For performing the 
continuum extrapolation, the results from three different lattice spacings corresponding to $N_\tau=8,10,12$ were considered. The $\chi_t$ shows 
a very characteristic temperature dependence and there is a change in the slope of the continuum curve at around $1.5 ~T_c$.  The temperature 
dependence is parameterized as $\chi_t^{1/4}=A T^b$ where $b$ determines the slope of the curve. This slope turns out to be $b=1.469(73)$ ~\cite{peter}
for temperature range between $1.1$-$2~ T_c$. We specifically compare the results for $\chi_t$ in this temperature interval measured on 
QCD ensembles generated from different staggered discretizations, HISQ~\cite{peter} vs stout~\cite{bonati} and domain wall fermions~\cite{viktorqm}. There is a very good agreement between 
results obtained with different lattice discretizations which strongly suggests for a more detailed study to explain such a temperature 
dependence immediately above $T_c$.

\begin{figure}[h]
\begin{center}
\includegraphics[width=0.45\textwidth]{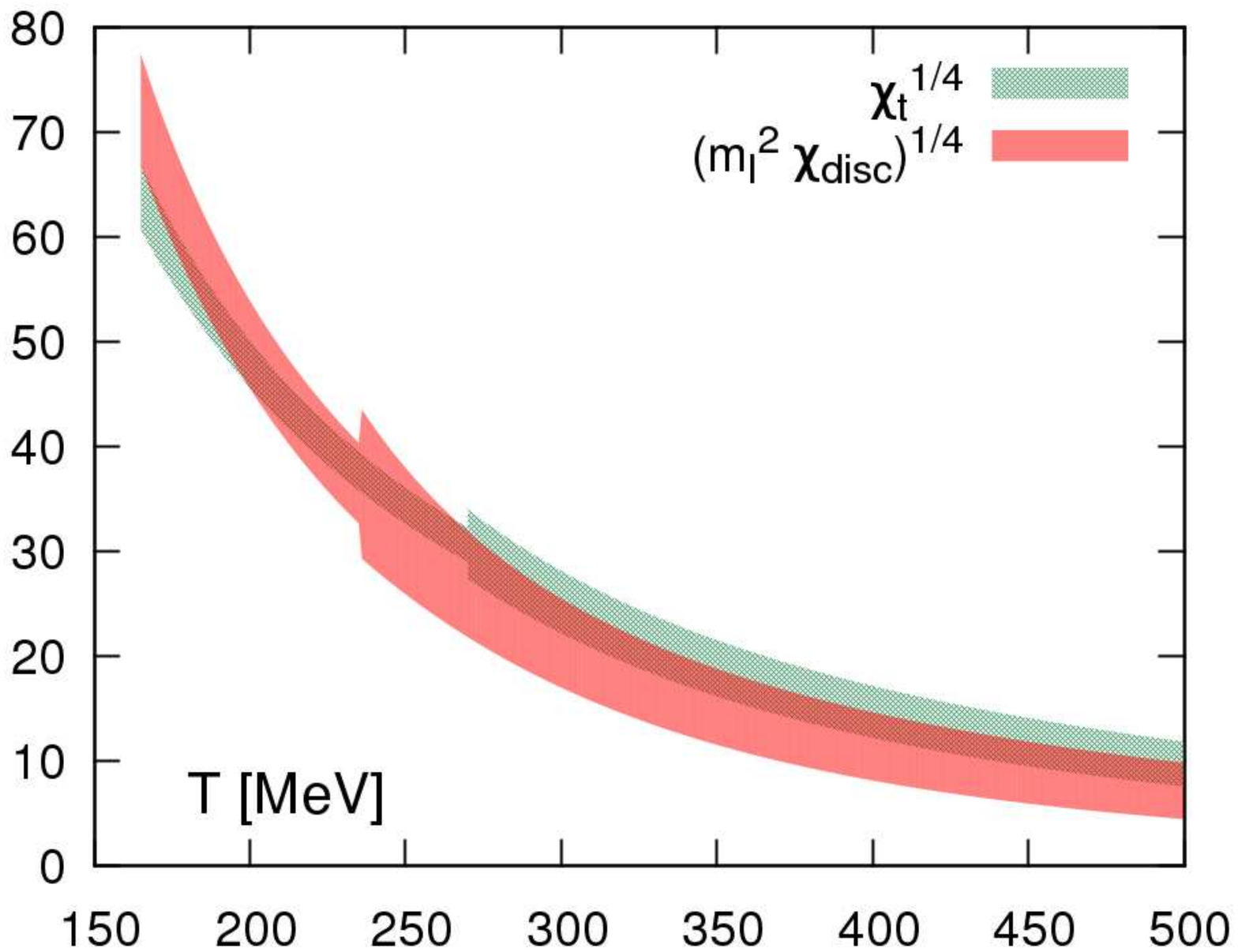}
\includegraphics[width=0.45\textwidth]{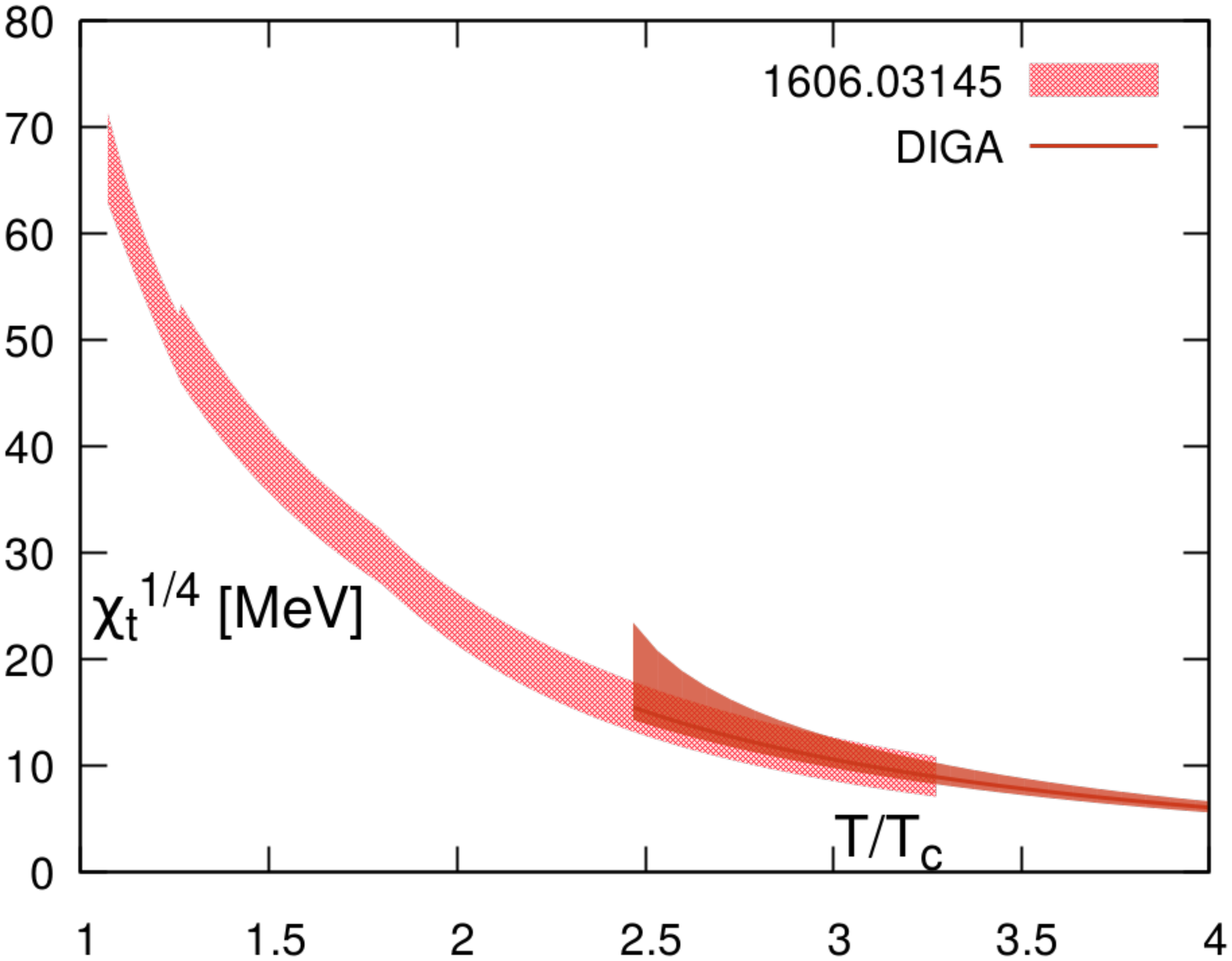}
\caption{The topological susceptibility measured with the gluonic definition is shown to agree with $m^2\chi_{disc}$ for a wide range in temperature in the 
left panel. In the right panel the continuum extrapolated results for $\chi_t$ is compared with the dilute instanton gas calculations. Though the temperature 
dependence of $\chi_t^{1/4}$ matches perfectly with the model predictions, the amplitude has to be scaled by a factor of $1.9$ to agree with dilute instanton gas 
model predictions which is known only upto leading order in semi-classical expansion.}
\label{fig:topsusc2}
\end{center}
\end{figure}

At higher temperatures beyond $2~ T_c$ the slope of $\chi_t$ as a function of temperature changes. Since cut-off effects become more severe at higher temperatures 
and topological tunnelings become rarer, it is important to ascertain whether our continuum extrapolation was performed correctly. 
Chiral Ward identities ensure that $\chi_t=m^2 \chi_{disc}$, $\chi_{disc}$ being the disconnected susceptibility which is a purely fermionic 
observable. Hence if one can verify this Ward identity it would ensure that the continuum extrapolation was done properly. Indeed 
the continuum estimates of these two quantities shown in the left panel of Figure \ref{fig:topsusc2} agree quite well for a wide
range of temperatures upto $\sim 4~T_c$. The exponent $b=1.85(15)$ characterizing the temperature dependence of $\chi_t$ beyond $2~T_c$ 
is in agreement with dilute instanton gas prediction $b\sim2$ for $N_f=3$ QCD. This is consistent with the conclusions from subsequent studies~\cite{borsanyi,mpl}. 
However the dilute instanton gas model calculation of $\chi_t$ is known only upto leading order in semi-classical expansion and second order in $\alpha_s$. The amplitude $A$ 
from the fit results of $\chi_t^{1/4}$ thus differs from the LO dilute gas prediction by a scale factor of $K=1.9$ at temperatures of $450$ MeV as shown in the right panel of Figure \ref{fig:topsusc2}. 
Knowing that the dilute instanton gas scenario sets in already at around $2~ T_c$ and there is no new underlying physics 
that would alter this behavior at higher temperatures, one can calculate the axion mass or alternatively the decay constant. If indeed the total 
dark matter density is due to QCD axions, the decay constant $f_a$ comes out to be of the order of $\sim 10^{12}$ GeV which is well within the parameter range which is probed 
by the ADMX experiment. One may wonder what is the impact of the uncertainty in the determination of this scale factor $K$ on this prediction. 
Varying the scale $K$ by $2 \sigma$ affects the prediction of $f_a$ by $\sim 15\%$ which is well within the experimental uncertainties~\cite{peter}.

\section{Conclusions and outlook}
In this section we summarize by formulating answers to the questions raised in the introductory section with the evidences collected from the studies so far. 
The infrared part of the eigenvalue spectrum of the QCD Dirac operator has very distinct features which are quite robust as we approach to finer lattice spacings 
independent of the fermion discretization we choose. The eigenvalue spectra consists of non-analytic peak and an analytic rise parameterized by $\vert\lambda\vert^\gamma$. 
The exponent $\gamma$ has a very strong temperature dependence above the chiral crossover transition. It changes from unity near $T_c$ by a factor of two at $1.2 ~T_c$ which is quite robust 
irrespective of the lattice discretization used. Both analytic and the non-analytic parts of the eigenvalue spectrum contribute to the breaking of $U_A(1)$ upto $1.5~ T_c$, 
through their contribution to observables like $\chi_\pi-\chi_\delta$. How large is the contribution of this non-analytic part and whether it survives in the continuum limit is 
still debated~\cite{jlqcd171}. However the eigenvalue spectrum of improved staggered quarks (HISQ) with finer lattices close to the continuum show evidence for the survival of the non-analytic 
peak which suggest that it may not be a lattice artifact and need much more further study. It has been argued earlier that at $T\sim1.5~T_c$ the existence of this peak can be understood arising 
from a non-interacting instanton-antiinstanton pair~\cite{viktor}. Further evidence from the study of temperature dependence of the topological susceptibility in QCD suggest that indeed dilute 
instanton gas scenario sets in early around $2~T_c$. This has exciting consequences since now it is possible to put in stronger constraints on the axion mass if indeed QCD axion is a 
dark matter candidate~\cite{peter,borsanyi}.

\section{Acknowledgments} 
This work has been supported through the contract DE-SC0012704 with the U.S. Department of Energy, Office of Science. Numerical calculations have 
been performed using the GPU cluster at Bielefeld University and the USQCD cluster at Jefferson Lab. The GPU codes used in our work were in part based on some publicly 
available QUDA libraries~\cite{quda}.


\begin{thebibliography}{99}
\bibitem{pw} 
R.\ D.\ Pisarski and F.\ Wilczek,  Phys.\ Rev.\  D 29, 338 (1984).
\bibitem{bpv}
A.\ Butti, A.\ Pelissetto, E.\ Vicari, JHEP 0308, 029 (2003); A.\ Pelissetto, E.\ Vicari, Phys. Rev. D 88, 105018 (2013); M.\ Grahl and D.\ H.\ Rischke, 
Phys.\ Rev.\ D 88,  056014 (2013); T.\ Sato and N.\ Yamada, Phys.\ Rev.\ D 91, 034025 (2015).
\bibitem{naka}
Y.\ Nakayama and T.\ Ohtsuki, Phys.\ Rev.\ D 91, 021901 (2015).
\bibitem{shailesh}
S. Chandrasekharan, N. Christ, Nucl. Phys.  Proc.Suppl. 47, 527 (1996).
\bibitem{dw12}
A. Bazavov et. al., Phys. Rev. D 86, 094503 (2012).
\bibitem{cossu}
G. Cossu et. al., Phys. Rev. D 87, 114514 (2013), Erratum: Phys. Rev. D 88, 019901 (2013).
\bibitem{hiroshi}
H.\ Ohno, U.\ Heller, F.\ Karsch, S.\ Mukherjee, Pos LATTICE2012 (2012), 095.
\bibitem{dw14}
M. I. Buchoff et. al., Phys. Rev. D 89, 054514 (2014); T.\ Bhattacharya et. al.,  Phys. Rev. Lett. 113, 082001 (2014).
\bibitem{viktor}
V.\ Dick et. al., Phys. Rev. D 91, 094504 (2015).
\bibitem{viktorqm}
S.\ Sharma et. al., Nucl. Phys. A 956, 793 (2016).
\bibitem{brandt}
B. Brandt et. al., JHEP 1612, 158 (2016).
\bibitem{jlqcd17}
A. Tomiya et. al., Phys. Rev. D96, 034509 (2017).
\bibitem{jlqcd171}
K. Suzuki et. al., arXiv:1711.09239 [hep-lat].
\bibitem{hotqcdeos}
A. Bazavov et. al., Phys. Rev. D 95, 054504 (2017).
\bibitem{ks}
T.\ Kalkreuter and H.\ Simma, Comput. Phys. Commun. 93, 33 (1996).
\bibitem{hotqcdeos1}
A. Bazavov et. al., Phys. Rev. D 85, 054503 (2012).
\bibitem{aoki}
  S.\ Aoki, H.\ Fukaya, Y.\ Taniguchi, Phys.\ Rev.\ D 86, 114512 (2012).
\bibitem{cossu2}
 G. Cossu et. al., Phys. Rev. D 93, 034507 (2016).
\bibitem{shuryak}
E.\ Shuryak,  Comments. Nucl.Part.Phys. 21 (1994)  235. 
\bibitem{yama}
T.\ Kanazawa and N.\ Yamamoto, Phys. Rev. D 91,  105015 (2015).
\bibitem{ivan}
A.\ Alexandru and I.\ Horvath, Phys. Rev. D 92, 045038 (2015).
\bibitem{peter}
P. Petreczky, H-P Schadler, S. Sharma, Phys. Lett. B 762, 498 (2016).
\bibitem{bonati}
C. Bonati et. al, JHEP 1603, 155 (2016).
\bibitem{borsanyi}
S. Borsanyi et. al, Nature 539, 7627, 69 (2016).
\bibitem{mpl}
F. Burger et. al., Nucl. Phys. A 967, 880 (2017).
\bibitem{quda}
 M.\ A.\ Clark et. al., Comput.\ Phys.\ Commun.\ 181, 1517 (2010).
 
 
\end{thebibliography}
\end{document}